# Water-induced self-oscillatory processes in colloidal systems by the example of instant coffee

**Tatiana A. Yakhno\* and Vladimir G. Yakhno**

Federal Research Center Institute of Applied Physics of the Russian Academy of Sciences (IAP RAS), 46 Ulyanov Street, 603950 Nizhny Novgorod, Russia

\*Correspondence to: yakhta13@gmail.com

Phone/Fax: +7(831) 436-85-80; mobile: +7(910) 384-39-44

Vladimir Yakhno: yakhno@appl.sci-nnov.ru

Phone/Fax: +7(831) 436-85-80

**Abstract**

Slow long-lived fluctuations of mechanical properties of drying drops of instant coffee water solutions have been found by means of the Drying Drop Technology developed earlier. Parameters of the fluctuations depended on the extent of dilution and were not affected by additional hashing of the solution, closing by a cover and shielding from external electromagnetic fields [1-3]. Here we compared dynamics of physicochemical and morphological processes in bulk. We observed microscopically periodic emergence, growth, aggregation and disappearance of spherical cavities (50-250 μm in diameter) in bulk of dispersion, which was followed by agreed fluctuations of physicochemical parameters of the colloidal system. We have shown that those spheres were huge hydrate covers of liquid crystal water around colloidal particles. Phase transitions between free and bound (liquid crystal) water in a bulk of a colloidal system is a pacemaker of fluctuations of its physicochemical properties. These changes are regulated and coordinated by the value of osmotic pressure. At low osmotic pressure hydrate covers grow, forcing out particles and ions into the bulk; at high osmotic pressure they collapse to small size, reducing osmotic pressure. Existence of self-oscillatory processes was confirmed by a mathematical model.

*Abbreviations*
DDT – Drying Drop Technology;
EZ – Exclusion zone;
SDS - sodium dodecyl sulfate

*Introduction*

As it often happens, our finding was casual. We tried to detect the origin of wide dispersion of data obtained at repeated testing of the same liquids on our device which measured dynamics of complex mechanical properties of drying drops – DDT [4-6]. It turned out that our repeated measurements had accurate temporal fluctuations [1-3]. We observed similar fluctuations in tea, wines, and coffee. Parameters of fluctuations depended on solute concentration and didn't depend explicitly on mixing, shielding from external electromagnetic fields and light [2, 3]. Such fluctuations of optical density were earlier observed in plasma and blood [7]. It seems that we deal with a universal phenomenon definitely of important fundamental and applied relevance for technique, biology and medicine [8].

Slow fluctuations in liquid media were also registered by means of other instrumental methods. Such fluctuations of light scattering in water, proteins and salts solutions were earlier noted by Chernikov [9, 10]. A detailed light scattering study showed that some regions in bulk can be characterized as close-to-spherical discrete domains of higher solute density in a less dense rest of solution [11]. These domains do contain solvent inside and can therefore be characterized as loose associates (giant clusters, aggregates). But this phenomenon has not yet found convincing explanation. At the end of the last century, a group of Japanese researchers reported results of their morphological observations [12]. Highly purified polymer latex dispersions had been studied with a confocal laser scanning microscope. In such dispersions, which were initially homogeneous, voids grew with time when dispersions were kept standing and formed more quickly in the internal material than in the material close to the glass-dispersion interface. Those events were reversible: after a while, the voids disappeared and the suspension became homogeneous forming a "colloidal crystal". Such an inhomogeneous distribution occurred, according to the authors, thanks to an electrostatic counterion – mediated attractive interaction between similarly charged colloidal particles [13-15]. It was shown that, generally, the spacing between the particles became smaller with increasing salt concentration. But when some threshold value of salt concentration was exceeded, the colloidal crystal melted [13]. The motion of a particle trapped inside a void in a colloidal dispersion was found to be highly restricted compared to the particles in the surrounding high-density region [16]. The authors hypothesized that the restricted motion was the result of a very long-range interaction with the surrounding particles, which had been overlooked in the colloidal interaction theory.

When studying dynamic processes in water environments, it is necessary to consider the phenomenon of formation and destruction of multilayered liquid crystal water at hydrophilic surfaces – exclusion zone (EZ) [17]. Colloidal and molecular solutes suspended in aqueous solution are profoundly and extensively excluded from the vicinity of various hydrophilic surfaces [18], so their concentration in bulk increases. The width of the solute-free zone is typically several hundred microns. Such large EZs were observed in the vicinity of many types of surfaces including artificial and natural hydrogels, biological tissues, hydrophilic polymers, monolayers, and ion-exchange beads, as well as in a variety of solutes.

Thus, the goal of this study was to disclose the mechanism of autonomous oscillatory processes in colloidal systems. From our preliminary studies and data available in the literature, it was clear that the causes of these processes should be sought in liquids.

*Materials and methods*

Our experiments were carried out using water solutions of the Nescafe Gold sublimated instant coffee bought in a store, with a concentration of 2.50 g/100 ml, at T = 22-23° C and H = 64-65%. A dry coffee sample was placed in a chemical glass, filled in with hot tap boiled water, and mixed by a glass stick until the coffee dissolved. Sampling was begun after cooling of solution to room temperature, at 9 o'clock Moscow time. Sampling was made each 30 minutes from the same glass of coffee solution standing on a table, using the zone equidistant from the center and edges of the glass, from the depth of about 2 cm by a microdispenser with removable tips. Such 30 minute intervals were stipulated by the duration of one test (20 min) and quartz treatment procedure. In some experiments we added to the colloidal system surfactant - sodium dodecyl sulfate (SDS) with a concentration of 0.2% w. Ten repeated experiments were made for every type of investigation.

*Drying Drop Technology (DDT)*

To monitor fluctuations of physicochemical properties of colloidal systems we used DDT based on acoustical impedancemetry developed in our laboratory earlier [1-3]. Here we explain only its main features. A coffee drop (volume of 3 μl) without any pretreatment dries on a polished end of a quartz plate. The quartz oscillates with a constant frequency of 60 kHz, which is equal to the resonance frequency of unloaded resonator. Acoustical – Mechanical Impedance (AMI) of the drop during drying is displayed as a curve on a screen (Fig. 1).

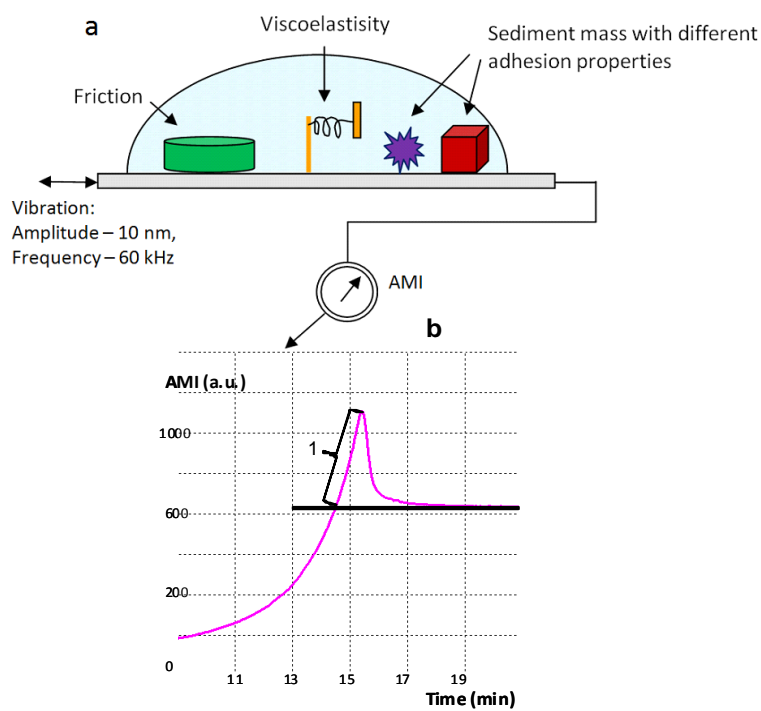

**Fig. 1**. DDT scheme: **a** – a drop as an object having a set of unique physical properties is drying on the surface of oscillating quartz resonator; **b** – typical AMI curve for drying drop of coffee: 1 – portion of the curve measured by total derivative (parameter BS_2).

The parameter BS_2 reflects the dynamics of complex mechanical properties of the drying drop deposit (mechanical stress) and is calculated automatically by software. In the same environment this parameter depends strongly on liquid composition and structure. We built up diagrams for temporal fluctuations of BS_2 parameter using Excel.

*Optical investigations of dried drops (coffee ring width detection)*

The tests were carried out simultaneously with BS_2 parameter measurements. For every 30-minute counting we took four drops having a volume of 3 µl: one drop for BS_2 parameter measurement, and 3 drops for coffee ring width measurements. Those 3 drops were placed on a new (without any treatment) microscope slides ApexLab, 7 countings of each slide (Fig. 2), up to 22 countings.

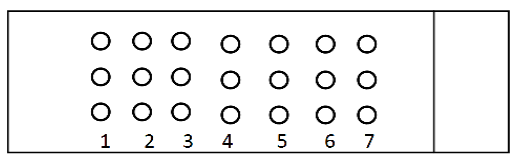

**Fig. 2**. Arrangement of drops on a glass for drying and microscopy.

The preparations were drying in horizontal position under room conditions, and were investigated the next day. We measured coffee ring width using the Levenhuk ToupView program in 3 positions into every drop (Fig. 3), so for each 30-minute account 9 measurements were made. Arithmetic mean and standard deviation were calculated for further analysis.

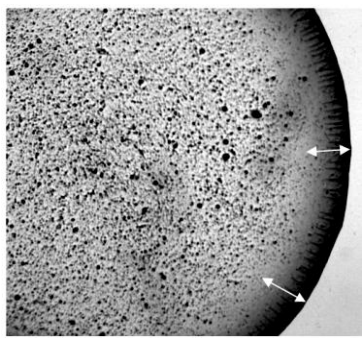

**Fig. 3**. Measurement of coffee ring width under microscope.

*Optical investigations of liquid samples*

The microscopy investigation of coffee solution was carried out in freshly prepared samples by the method of "flattened drop". For this purpose a drop with a volume of 5 µl was placed on a new (without any treatment) microscope slide ApexLab (25.4 x 76.2 mm) then the drop was covered with a cover glass 24 x 24 mm in size (ApexLab), avoiding formation of air bubbles, and was studied under microscope Levenhuk with a digital camera connected to a computer. We made 10 pictures for every 30-minute step with the same magnitude and analyzed them later using the Levenhuk ToupView program. Morphometric measurements (diameters of avoids in the pictures) were made for every 30-minute step. Statistical analysis (calculation of mean and standard deviation) were made by Excel program. In some experiments, in parallel with the flattened drop, freshly prepared smears (without cover glass) were also examined under a microscope in polarized light.

*Surface tension fluctuations detection*

For detecting surface tension temporal changers we used a set of certified glass capillaries (10 μl Drummond Microdispenser, 100 Replacement Tubes, Made in the USA by Drummond Scientific Company. Cat.# 3-000-210G). Each capillary was used once. A new dry capillary was submerged into liquid at regular intervals to a certain mark on a capillary and liquid raising level was measured. Simultaneously, the fluctuations of BS_2 were usually measured. Diagrams and calculations of the correlation coefficient were done by means of Excel program.

*Results and Discussion*
*1. DDT Investigations*

Figure 4 shows joint temporal fluctuations of parameter BS_2 and coffee ring width. Direct linear correlation between them at significant value p = 0.005 was 0.7 ± 0.16. This testified to a causal relationship between these parameters.

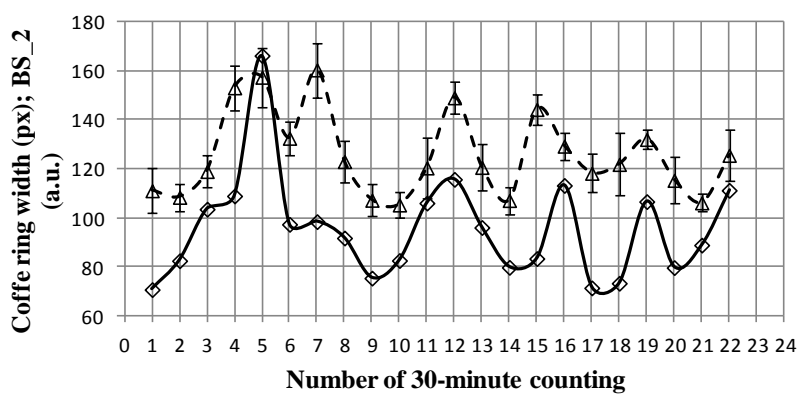

**Fig. 4.** Joint dynamics of parameter BS_2 (solid line) and coffee ring width (dashed line) in coffee water solution (2.5 g/100 ml).

Fluctuations of surface tension in the same coffee solution could be measured more frequently. It was shown that one period took 30-40 minutes (Fig. 5).

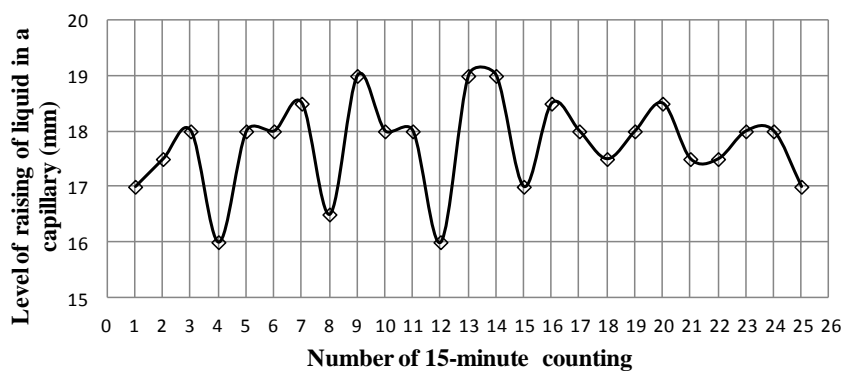

**Fig. 5.** Fluctuations of surface tension in coffee water solution (2.5 g/100 ml).

Correlation coefficient between BS_2 and surface tension fluctuations in one and the same experiment was 0.8 ± 0.2 (p = 0.01) (Fig. 6).

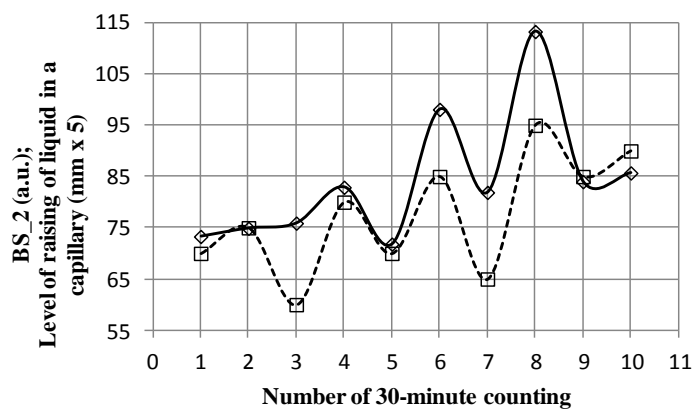

**Fig. 6.** Joint dynamics of parameter BS_2 (solid line) and level of rising of liquid in a capillary (dashed line) in coffee water solution (2.5 g/100 ml).

It is interesting to note that fluctuations of these parameters didn't disappear either on the third day of stay of this liquid in the same glass without cover on the table in laboratory. Despite a long period of storage, fluctuations of parameters persist, and direct correlation link between them remains high (r = 0.7 ± 0.2, p = 0.01).

It was important to find out how addition of surfactant influences parameters of fluctuations and structurization of the drops drying on a glass support. According to Fig. 7, SDS adding drastically reduced BS_2 value (mechanical stress during drop drying). It occurred due to decrease in interaction between colloidal particles as well as between the particles and quartz surface. Diameter of the dried drops considerably increased, and the relief of their surfaces became smooth (Fig. 7, drops 12, 13). Drops 5 and 6, corresponding to one of maxima of fluctuations of the BS_2 parameter before SDS adding were characterized by the relief coffee ring and presence of fragments of reticular structures on the surface. Drops 4 and 7, corresponding to the minimum BS_2 values, had more flat coffee ring and didn't contain the reticular structures. Instead of them separate clamps on a surface of the drops were observed.

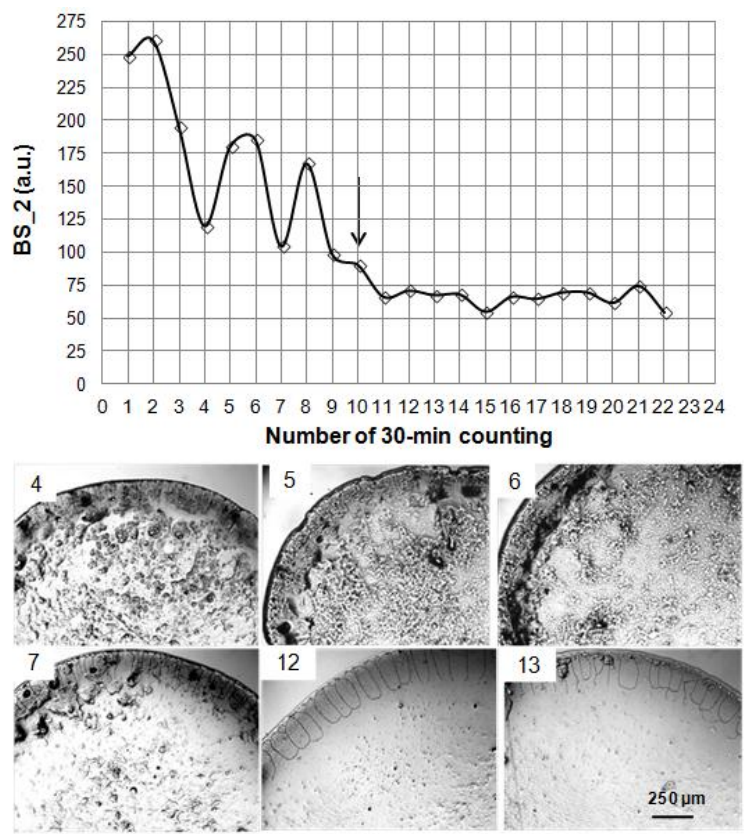

**Fig. 7.** Fluctuations of the BS_2 parameter in the drying drops of coffee solution (2.5 g / 100 ml) before and after SDS addition (the moment of addition is specified by an arrow). From below - photos of the dried drops of coffee solution on glass support taken from total volume in different phases of the process (numbers of photos correspond to numbers of counting in above diagram).

Our data agree with results of the work [19], showed that the interactions of colloids with (and at) liquid-solid and liquid-gas interfaces as well as bulk particle-particle interactions affect the morphology of the deposit. Now we can add that such interactions influence also the mechanical properties of dried materials from these colloids, which may be represented quantitatively. After surfactant addition the area of drops considerably increased, formation of the coffee ring has been complicated and structurization was suppressed, which corresponds to results of the research [20].

Thus, we can state that autonomous temporal fluctuations of mechanical properties of drying drops of colloidal suspensions revealed by us earlier [1-3], are also followed by coordinated fluctuations of surface tension. We will try to disclose the internal mechanism of these fluctuations looking directly into a liquid phase.

*2. Morphological investigations of liquid phase*

Observation of colloidal systems under optical microscope followed the same scheme: we took samples from one and the same volume of coffee solution in certain periods of time and investigated them by the method of flattened drop.

Perfectly shaped circles contoured by colloidal particles, sitting close to each other were observed everywhere (Fig. 8). The circles were sitting on a glass substrate. Commonly it could be possible to find one central particle in each circle. Those round figures could associate, forming large – scale agglomerates.

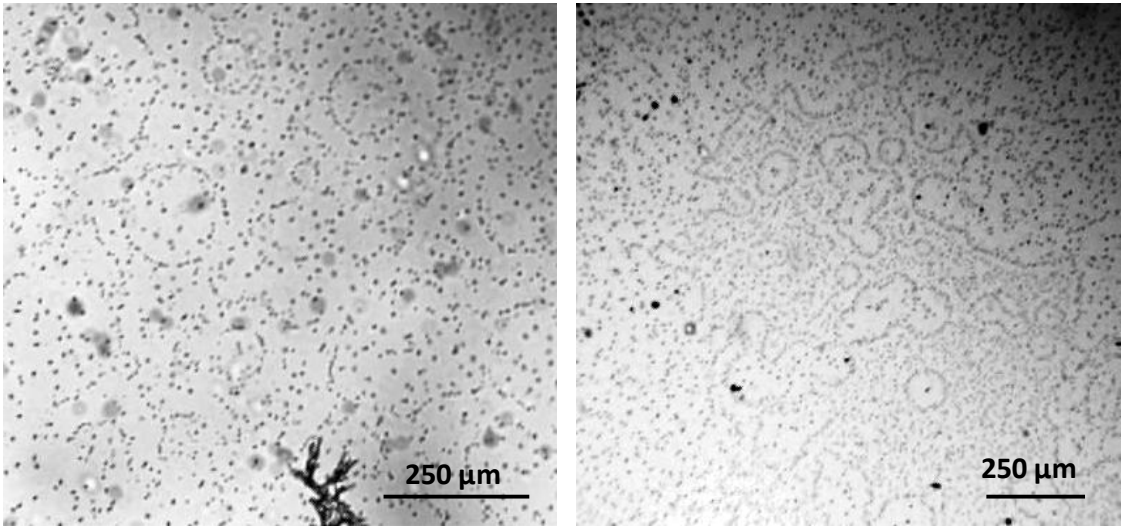

**Fig. 8.** Microphoto of water solution of coffee. Round figures and associates of round figures.

It seems that growing "circles" pushed back colloidal particles, creating conditions for their convergence and coagulation. The size of the particles observed by us was not less than 1 μm so they did not participate actively in Brownian motion. Therefore during creation of spatial reticular structures their passive crowding due to the growing external structures seems to us more convincing than their active movement at the expense of the long-range attraction forces. Figure 9 shows stages of temporal evolution of round structures in bulk, from small to big, and the remains of arches from colloidal particles after collapse of "round structures".

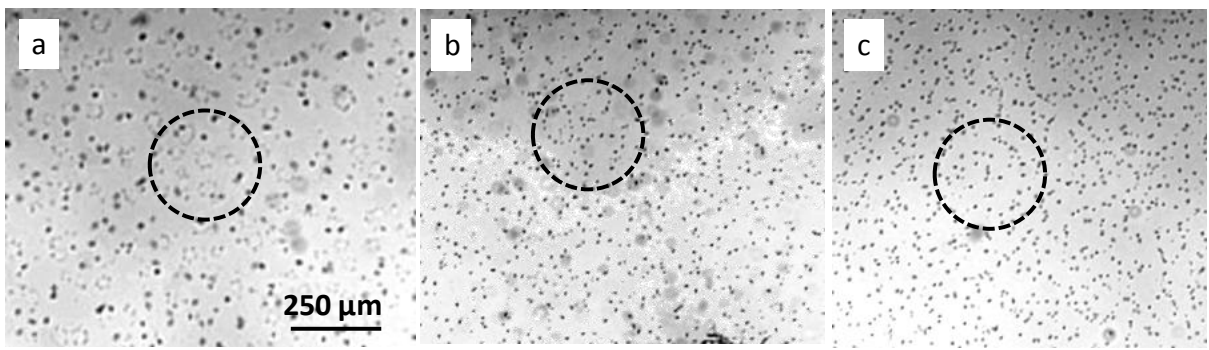

**Fig. 9.** Microphoto of water solution of coffee. Temporal evolution of round structures in bulk. Some structures in every picture are encircled (as a guide for eyes). Sampling time from the solution: a – 12:50, b – 13:30, c – 14:00.

Similar arches after collapse of round structures could be observed for some time in free floating (Fig. 10).

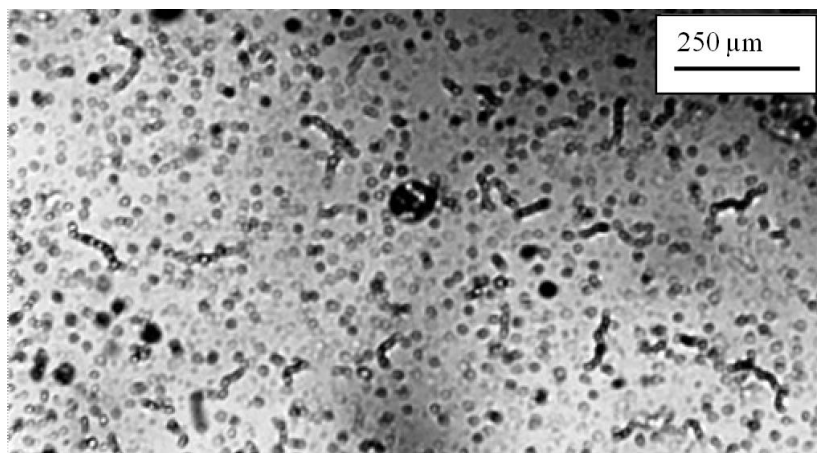

**Fig. 10.** Water immersion. Microphoto of water solution of coffee. Remainders of arches floating in solution after destruction of round structures.

The dynamics of growth and destruction of such round structures and their associates is shown in Fig. 11.

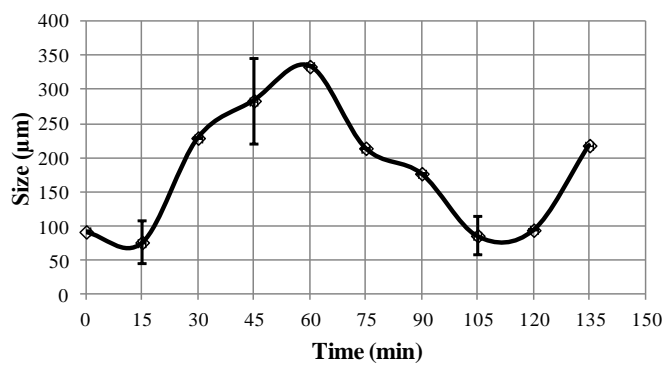

**Fig. 11.** Dynamics of growth and destruction of "round structures" in coffee solution (2.5 g / 100 ml).

On the ascending and descending parts of the curve, size distribution of structures became bimodal due to the presence in the field of view, along with round structures, their large associates (see Fig. 8, left). Nevertheless, our observations have revealed the rhythmic nature of formation and destruction of round structures, similar to a rhythm of fluctuations of physicochemical parameters of this colloidal system. Our equipment allowed observing events only in two-dimensional option. Therefore, circular structures can be a projection of the balls on the plane. Data obtained by means of a laser scanning microscope manifestly showed spherical cavities in latex suspension (Fig.2 in [12]). Unfortunately, the authors did not pay attention to their shape. Those cavities looked empty, but now we believe that they were filled by transparent liquid crystal water. If so, then it is easy to explain the restricted movement of the particle placed in such media [16]. This assumption is confirmed by our observations of freshly prepared smears of coffee solution (Fig. 12, c-f). We could see real agglomerates of liquid crystal water, which started melting after a while. In a flat variant (between substrate and cover glasses), these agglomerates consist of round structures, which have visible borders due to adsorbing colloid particles (Fig. 12, a,b).

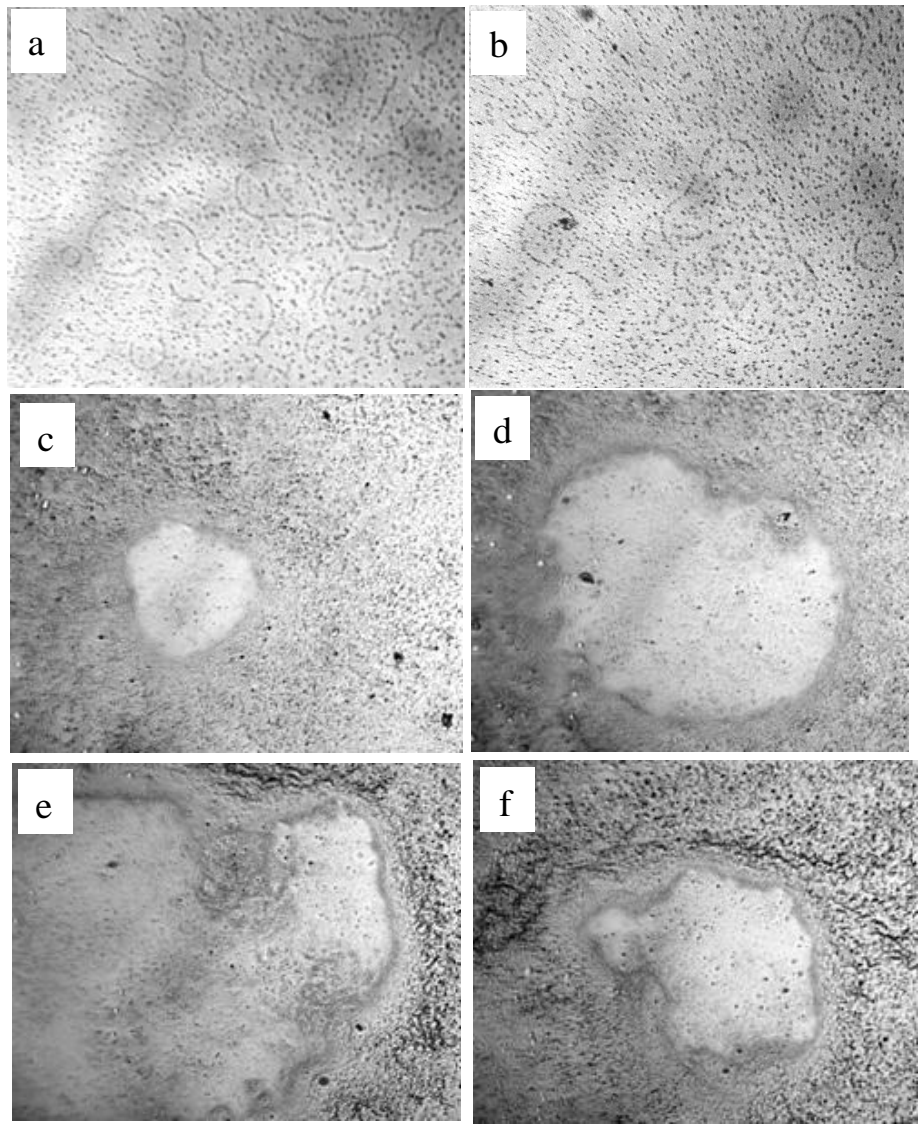

**Fig. 12.** Microphoto of water solution of coffee (a,b) prepared by the method of flattened drop; c-f - smears of the same coffee solution in polarized light. Field width – 1700 μm. c, d – crystal water agglomerates just after preparation; e, f – crystal water agglomerates after a while (melting process).

The mechanism of particle interaction in solution is currently actively discussed. Attraction of like charged gel beads with a diameter of 400-650 μm spaced several hundred micrometers apart in water was described in [21]. The authors measured the charge distribution around the beads with a pH sensitive dye and conjectured that the cause of the long-range attraction was a shell of multilayer structured water, formed around beads' hydrophilic surface. Here we can see the analogy with our experiment, where colloidal particles rather than gel beads interact. Moreover, their interaction is caused by spatial convergence which is due to the growing spheres of the liquid crystal water. In soft matter and nano-science, critical Casimir forces attract an increasing interest thanks to their capability of reversible particle assembly [22-28]. These forces are the thermodynamic analogue of the quantum mechanical Casimir force arising from the confinement of vacuum fluctuations of electromagnetic field. In its thermodynamic analogue, solvent fluctuations confined between suspended particles give rise to an attractive or repulsive force between them. Due to its unique temperature dependence, this effect allows in situ control of reversible assembly [26, 27]. The authors of [23] showed that in the system with negligible van der Waals forces a simple competition between repulsive screened Coulomb and attractive critical Casimir forces can be accounted quantitatively for the reversible aggregation. Above the temperature T$a$, the critical Casimir force drives aggregation of the particles into fractal clusters, while below T$a$, the electrostatic repulsion between the particles breaks up the clusters, and the particles resuspend by thermal diffusion [23]. If the gap between the interacting surfaces is filled with a specially designed substance, the attraction between the surfaces can change their repulsion. If such interaction of surfaces with a dielectric constant $\mathcal{E}_1$ or $\mathcal{E}_2$, respectively, occurs in a medium with a dielectric constant $\mathcal{E}_3$, they will be attractive at $(\mathcal{E}_1 - \mathcal{E}_3)(\mathcal{E}_2 - \mathcal{E}_3) < 0$, and repulsive at $(\mathcal{E}_1 - \mathcal{E}_3)(\mathcal{E}_2 - \mathcal{E}_3) > 0$. These interactions are extremely sensitive to temperature, chemical composition of the medium and its physical characteristics [24,25].

According to our data, the observed process is characterized by cyclic changes both in liquid solute concentration due to displacement of the ions and particles from EZs to the bulk, and in particle surface properties due to EZ shell growth around them. As these zones routinely generate protons in the water regions beyond, unequal proton concentrations in the respective areas may be responsible for creating both the pH and potential gradients, which may be ultimately responsible for the osmotic drive [29]. On the other hand, the surface water has different water activity and chemical potential to the bulk, leading to differences in osmotic pressure and other colligative properties [30]. When this increase in osmotic pressure next to the surface reaches a threshold, the mechanical instability of the system sharply growths, velocity of microstreams enhances, and EZ spheres start to collapse. They break into small pieces and melt. Solute concentration and osmotic pressure decline. Free colloidal particles are distributed uniformly. Chains of particles coagulated on the surface of the water balls remain in solution. Growth of EZ balls begins again and the process recurs (Fig. 13).

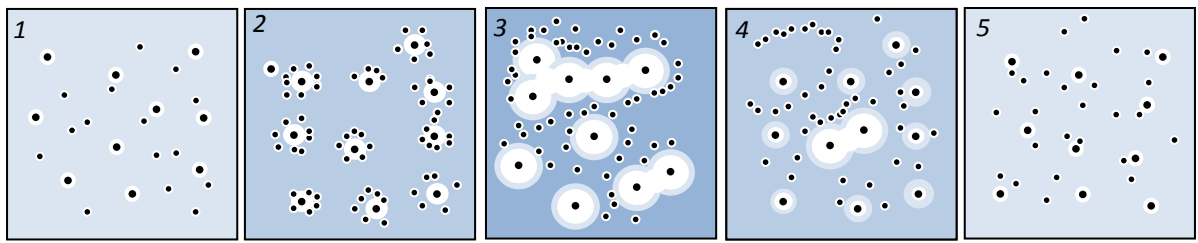

**Fig. 13.** Scheme of cyclic physicochemical transformations in colloidal system. Background coloring intensity corresponds to the concentration of ions and particles in dispersive phase. *1* – Initial state, quasi-homogenous distribution of particles with small EZs; *2* – EZs growth around some particles is more intense than around others; *3* – giant EZ-balls aggregate and begin destruction due to high osmotic pressure; chains of particles coagulated on the surface of water balls remain in solution; *4* – destruction process continues, osmotic pressure decreases progressively; *5* – EZs are ready to grow again.

As similar events (EZ growth) are registered for other polar liquids (10, 31), we believe that the autonomous fluctuations based on rhythmic formation and destruction of liquid crystal spheres are the universal law for colloids.

The considered processes have been used for creation of a phenomenological model showing a possibility of the existence of self-oscillatory modes in similar systems.

Let the volume of a colloidal system be a cube with edge length $L$. Let this cube house $N_1$ hydrophilic particles – seeds around which liquid crystal water spheres (EZ) are formed. Let $n$ be the number of ions and colloidal particles determining osmotic pressure $P$ at the interface between water spheres of radius $r$ and dispersive medium. $V$ is the amplitude of mean speed of microflows with a characteristic lateral dimension smaller than $\pi r$. This corresponds to the excited mode of mechanical instability for the sphere surface: $2\pi r/m$, where $m = 2, 3, 4, 5…$
The estimated equations for integral processes in such a system can be written down in the following form:
EZ growth near a seed particle can be described as

$$dr/dt = l_0/\tau_{gr}(1 - V/V_{crit}) \quad , \tag{1}$$

where $V$ is the average velocity of microstreams in bulk near the water balls;
$V_{crit}$ is the critical velocity of microstreams in bulk with sufficient energy for destruction of external borders of the water balls;
$l_0$ is the increment of EZ shell thickness around a hydrophilic particle during time $\tau_{gr}$.
From the work [18] and our own experiments we know that $l_0/\tau_{gr} \approx 1 - 10$ µm/sec.
Formation of microstreams near the interface between the EZ shell and free water on achieving critical osmotic pressure $P_{crit}$ can be described as

$$dV/dt = -V/\tau_{visc} + 4\pi r^2 \cdot \gamma_m \cdot P \cdot F_{[x]} \cdot [P - P_{crit}] \quad , \tag{2}$$

where $\tau_{visc}$ is the characteristic time of reduction of microstreams velocity due to solution viscosity; $\tau_{visc} \approx$ const;
$\gamma_m$ is the coefficient characterizing average change of destruction force depending on the created mode of spatial nonuniformity on the destroyed external border of EZ;
$F_{[x]}$ is the step-type function equal to zero if $x = P - P_{crit} < 0$, and equal to 1 if $x = P - P_{crit} > 0$.

We assume that the speed of diffusion of ions and colloidal particles is much more than the growth rate of EZ shell and speeds of delay of microstreams. Then we can use Vant Hoff's law for stationary conditions

$$P = n \cdot R \cdot T/V \quad , \tag{3}$$

where $R$ is universal gas constant, $T$ is absolute temperature, and V is a volume of colloidal liquid.

The volume of colloidal liquid except for the volume of water spheres is found from equation

$$V = L^3 \cdot [1 - 4/3\pi r^3 \cdot N_1/L^3] \tag{4}$$

Thus, the status of water spheres in the bulk of the remaining colloidal liquid can be described by equations (1-4). We introduce the following notation

$$\beta = 4\pi N_1 / 3L^3 ; \tag{5}$$

$$\alpha = 4\pi \gamma_m \cdot nRT / L^3 \cdot V_{crit} ; \tag{6}$$

$$\chi = V / V_{crit} \tag{7}$$

and rewrite the above equations correspondingly:

$$\begin{cases} dr/dt = l_0 / \tau_{gr}(1-\chi) & (8) \\ \\ d\chi/dt = -\chi/\tau_{visc} + \alpha r^2 / 1 - \beta r^3 \cdot F_{[x]} \cdot [P - P_{crit}] & (9) \\ \\ P = n \cdot R \cdot T / L^3 \cdot (1 - \beta r^3) = (V_{crit} / 4\pi \gamma_m) \cdot \alpha / (1 - \beta r^3) & (10) \end{cases}$$

On the basis of this system of equations and understanding of the physics of the dynamical process we can distinguish 3 stages of the process:
1) EZ shell growth around hydrophilic colloidal particles to the size of huge liquid crystal water spheres; osmotic pressure growth in a bulk;
2) Osmotic pressure growth in bulk over critical value, forming conditions for the development of mechanical instability at the interface between water spheres and bulk (similar to the Rayleigh-Taylor instability [32]); microstream strengthening, causing erosion until complete destruction of water spheres.
3) Microstreams slowdown due to viscosity and transition of the system to stage 1.

Let's consider the dynamics of the process based on equations (8-10) in simplified form.

Stage 1. EZ shell growth around hydrophilic colloidal particles (11-13):

$$\begin{cases} dr/dt = l_0 / \tau_{gr}(1-\chi) & \chi < 1 & (11) \\ \\ d\chi/dt = -\chi/\tau_{visc} & \chi \approx 0 \quad (\text{at } t=0) & (12) \\ \\ P = (V_{crit} / 4\pi \gamma_m) \cdot \alpha / (1 - \beta r^3) < P_{crit} & (13) \end{cases}$$

Stage 2. Development of instability and destruction of water spheres (14-16):

$$P = P_{crit} \tag{14}$$

$$r^3 = r^3_{crit} = 1/\beta \cdot [1 - (\alpha \cdot V_{crit} / 4\pi\gamma_m \cdot P_{crit})] \tag{15}$$

$$d\chi/dt = (\chi_{max} - \chi) / \tau_{visc}, \tag{16}$$

where $\chi_{max} = \alpha \cdot \tau_{visc} \cdot r^2_{crit} / (1 - \beta r^3_{crit})$. If $\chi \leq 1$, then $r$ continues to grow. According to our observations, $r_{crit} \approx 250$ μm.

Stage 3. Microflow slowdown.

Since turbulent flows are formed at this stage, $\tau_{gr}$ may depend on $\chi$ and P. However, for our simplified representation, we'll assume that $\tau_{visc}$ = const, as in the case of laminar flow (17-20):

$$P < P_{crit}. \tag{17}$$

$$dr/dt = l_0/\tau_{gr} (1 - \chi) \qquad \chi < 1 \tag{18}$$

$$d\chi/dt = -\chi / \tau_{visc} \tag{19}$$

$$P = (V_{crit} / 4\pi\gamma_m) \cdot \alpha / (1 - \beta r^3) < P_{crit} \tag{20}$$

This dynamics can also be represented on phase plane in $\chi$ and $r$ coordinates (Fig. 14).

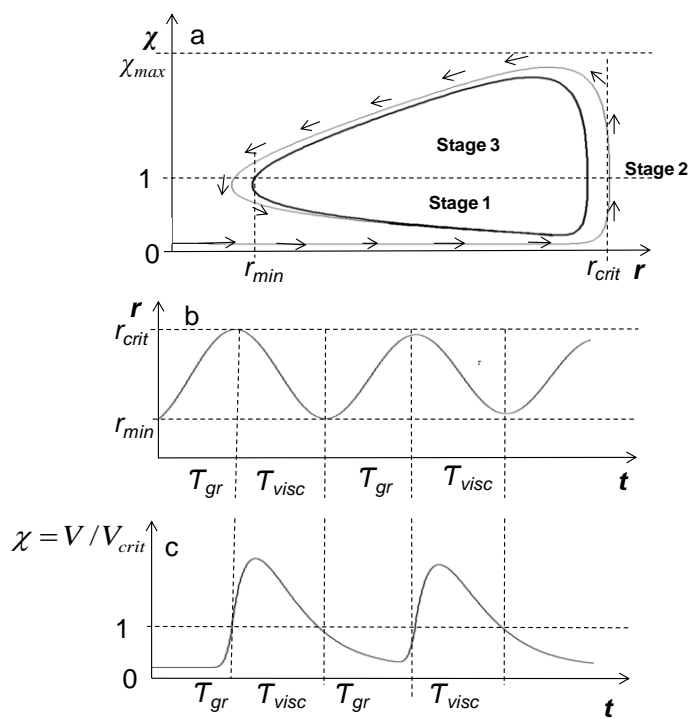

**Fig. 14.** Qualitative description of cyclic changes of variables in the course of growth of water spheres, their destruction and formation of conditions for their new growth: a - view of phase trajectories: process begins at the zero point ($\chi = 0; r = 0$) and reaches limit cycle; b - temporary change of radius of water spheres; c - temporary change of velocity of microstreams.

*Conclusions*

In this work we used instant coffee as a sample of complex colloidal system. Making long repeated measurements of the dynamics of complex mechanical properties of drying drops by DDT [4-6] we noticed slow periodic fluctuations of the measured parameter. Additional studies showed that, simultaneously with the above mentioned parameter, there occurred coordinated fluctuations of the surface tension and the character of structuring in drying drops. We had found earlier that parameters of fluctuations depended on the extent of dilution and did not depend on additional hashing of solution and shielding from external electromagnetic fields [2, 3]. So, we decided to elucidate the mechanism of such fluctuations in liquid phase. The basis for such a study was, on the one hand, information available on reversible formation of voids inside colloidal dispersions [12,16] and, on the other hand, description of slow fluctuations and inhomogenity of fluids of different types obtained in the light scattering studies [9-11]. Moreover, we had earlier described similar fluctuations of optical density in blood and plasma diluted with physiological saline solution in different proportions [7, 8]. For the liquid phase study, we used the method of flattened drop and freshly prepared smears of coffee solution. We described for the first time the existence of spherical structures in the liquid phase of coffee, their periodical occurrence, growth, destruction and re-emergence, which agreed with fluctuations of physicochemical parameters of the system. We showed morphologically that these spheres are liquid crystal water which forms EZs around hydrophilic colloidal particles. Agglomerates of such liquid crystal water spheres look like voids inside colloidal dispersions under confocal scanning laser microscope. Our findings are based on modern and last-century studies of wall layer of water mentioned in the book of Gerald Pollack [17]. We proposed a water-induced mechanism of self-oscillatory processes in colloidal systems and confirmed the principle possibility of its existence by a simple mathematical model.

These data allow a fresh look at the process of aggregation - disaggregation of colloidal particles in the solution. The experiments showed that the growth of water spheres pushes colloidal particles to the borders of the spheres, promotes their crowding and coagulation. On the border of water spheres, the particles form chains (reticular structures) which exist some time in liquid phase after destruction of liquid crystal water shells. Thus, all complex dynamics is controlled by the phase transitions of water – from the free to the bound (liquid crystal) state and back. Osmotic pressure acts as the intermediary messenger and the synchronizer of these transformations in a whole volume of liquid. Actually, these processes are not very sensitive to temperature (unlike the models based on calculation of Casimir forces [23, 26, 27]), and do not notably depend on liquid disturbance by hashing. Hashing by a turning of a test tube initiates emergence of streams of liquid with a characteristic size of an order of the size of a test tube (1.5 x 9.0 cm [2,3]), when the most part of the brought energy is spent for movement of spheres in bulk. Destruction of water spheres happens at initiation of microstreams of 10-100 μm in size, which correspond to the sizes of the destroyed objects. Osmotic pressure P in accordance with equation (3) is proportional to the absolute temperature ($\approx$ 295 K). Possible changing the temperature at a few degrees is negligible to affect the considered processes. For obvious reasons parameters of fluctuations depend on concentration of the components. The described mechanism can explain some phenomena that were not clear before and deserves special research. We believe that we deal with a universal phenomenon of undoubted importance for fundamental and applied science. We are sure that our hypothesis will stimulate researchers with other ideas and other tool kits to join this direction of study.

*Acknowledgements*

This research did not receive any specific grant from funding agencies in the public, commercial, or not-for-profit sectors. This work was done in the framework of state project no. 0035-2014-0008 (Institute of Applied Physics, Russian Academy of Sciences).

The authors are grateful to their colleague Anatoly Sanin for technical support of the work.